# Light bullets in Bessel optical lattices with spatially modulated nonlinearity

Fangwei Ye,[1] Yaroslav V. Kartashov,[2] Bambi Hu[1,3] and Lluis Torner[2]

[1]*Department of Physics, Centre for Nonlinear Studies, and The Beijing-Hong Kong Singapore Joint Centre for Nonlinear and Complex Systems (Hong Kong), Hong Kong Baptist University, Kowloon Tong, China*
[2]*ICFO- Institut de Ciencies Fotoniques, and Universitat Politecnica de Catalunya, Mediterranean Technology Park, 08860, Castelldefels (Barcelona), Spain*
[3]*Department of Physics, University of Houston, Houston, Texas 77204-5005, USA*
*Fwye@hkbu.edu.hk*

**Abstract:** We address the stability of light bullets supported by Bessel optical lattices with out-of-phase modulation of the linear and nonlinear refractive indices. We show that spatial modulation of the nonlinearity significantly modifies the shapes and stability domains of the light bullets. The addressed bullets can be stable, provided that the peak intensity does not exceed a critical value. We show that the width of the stability domain in terms of the propagation constant may be controlled by varying the nonlinearity modulation depth. In particular, we show that the maximum energy of the stable bullets grows with increasing nonlinearity modulation depth.

## References and links

Light bullets, or optical spatiotemporal solitons, are one of the most exciting entities in nonlinear optics. They are localized wavepackets in which both diffraction and dispersion are balanced by nonlinearity [1,2]. Being fully three-dimensional states, light bullets are unstable against collapse in uniform focusing Kerr media. A number of settings have been proposed where their stabilization is possible. Thus, stable bullets may form in materials with saturable or competing nonlinearities [3-5], quadratic [6,7] or nonlocal [8,9] nonlinear media, optical tandem geometries [10, 11], and also in waveguide arrays and optical lattices imprinted in different materials [12-17].

Typically, when considering light bullet formation in optical lattices the nonlinearity of the material is assumed to be homogeneous. However, current techniques allow engineering both, the linear refractive index and the nonlinearity of the material. This is the case, for example, in Bose-Einstein condensates (BECs) loaded in optical lattices, where spatial modulation of the interatomic interaction strength can be achieved by using Feshbach resonances by applying inhomogeneous external magnetic [18] or optical [19] fields. Such possibility has stimulated numerous theoretical studies of solitons in BECs with local modulation of the mean-field nonlinearity or even BECs in periodic nonlinear lattices [20-24]. Engineering the linear refractive index together with the nonlinearity may be also possible in optical structures, e.g., in photonic crystals with the holes infiltrated with a highly nonlinear material, such as suitable index-matching liquids [25-28]. Nonlinearity may also be modulated by changing the concentration of dopants upon fabrication. Such modulation has been demonstrated, e.g., in waveguide arrays made by Ti in-diffusion in Lithium Niobate crystals [29]. Analogously, nonlinearity modulation can be achieved in arrays written in glass by high-intensity femtosecond laser pulses where the optical damage produced by the tightly focused laser beams causes an increase of the refractive index accompanied by a simultaneous decrease of the nonlinearity coefficient [30,31].

The properties of one-dimensional (1D) and two-dimensional (2D) solitons in mixed linear-nonlinear lattices are known to depart substantially from the properties of conventional lattice solitons [32-41]. However, to date, the properties of fully three-dimensional light bullets in geometries with simultaneous modulation of the linear refractive index and the nonlinearity have not been addressed. Such investigation is interesting from a fundamental point of view, because examples of settings where stable three-dimensional solitons exist are relatively rare, and many of them also rely on complex composite geometries and materials. Moreover, since three-dimensional solitons in uniform cubic media undergo supercritical collapse, addition of a linear lattice results in their stabilization only in a certain limited range of parameters. Under such conditions an additional modulation of the cubic nonlinearity may dramatically affect the domains of existence and stability of the lattice-supported light bullets. Geometries with competing linear and nonlinear lattices, where the refractive index is decreased in the points where the nonlinearity peaks are particularly intriguing.

In this paper we address fully three-dimensional light bullets in a two-dimensional optical Bessel lattice with out-of-phase modulation of the linear refractive index and the nonlinearity. We find that the maximum energy at which light bullets remain stable increases with the depth of the nonlinearity modulation. Interestingly, we find that in this setting the so-called Vakhitov-Kolokolov (VK) criterion accurately predicts the domains of stability and instability *only* when the nonlinearity modulation is not too strong.

We describe the propagation of a light beam along the $\xi$ axis of a medium with imprinted out-of-phase linear and nonlinear lattices, by using the nonlinear Schrödinger equation for the dimensionless complex field amplitude $q$:

$$i\frac{\partial q}{\partial \xi} = -\frac{1}{2}\left(\frac{\partial^2 q}{\partial \eta^2} + \frac{\partial^2 q}{\partial \zeta^2}\right) + \frac{\beta}{2}\frac{\partial^2 q}{\partial \tau^2} - [1 - \sigma R(\eta,\zeta)]|q|^2 q - pR(\eta,\zeta)q. \qquad (1)$$

Here $q(\eta,\zeta,\tau,\xi) = (L_{\text{dif}}/L_{\text{nl}})^{1/2} A I_0^{-1/2}$; $A$ is the slowly varying envelope; $I_0$ is the input intensity; $\xi = z/L_{\text{dif}}$ is the propagation distance normalized to the diffraction length $L_{\text{dif}} = k_0 r_0^2$; $\eta = x/r_0$ and $\zeta = y/r_0$ are the transverse coordinates normalized to the characteristic transverse scale $r_0$; $\tau = t/t_0$ is the normalized (retarded) time; $t_0$ is the characteristic pulse duration; $L_{\text{nl}} = n_0/(k_0 n_2 I_0)$ is the nonlinear length; $\beta = L_{\text{dif}}/L_{\text{dis}}$; $L_{\text{dis}} = -t_0^2/(\partial^2 k_0/\partial \omega^2)$ is the dispersion length; $p$ and $\sigma$ describe the modulation depths of the linear refractive index and nonlinearity, respectively; $R(r) = J_0[(2b_{\text{lin}})^{1/2}r]$ stands for the profile of a radially symmetric lattice, with $r = (\eta^2 + \zeta^2)^{1/2}$ being the radius and $b_{\text{lin}}$ a parameter that characterizes the transverse lattice scale. For convenience, here we set $b_{\text{lin}} = 2$. Here we consider Bessel lattices, but it is worth noticing that the main conclusions are expected to hold for other similar types of radially symmetric lattices. Notice that the nonlinearity coefficient $1 - \sigma R(r)$ acquires local minima (maxima) in the points where the linear lattice $pR(r)$ attains maxima (minima). We refer to this type of modulation as "out-of-phase". We consider here only the case $\sigma \leq 1$ when the nonlinearity coefficient $1 - \sigma R(r)$ remains positive for any value of $r$. Equation (1) conserves the total energy flow $U$, Hamiltonian $H$ and $\xi$-projection of angular momentum $L_\xi$:

$$U = \iint \int_{-\infty}^{\infty} |q|^2 \, d\eta d\zeta d\tau,$$

$$H = \frac{1}{2} \iint \int_{-\infty}^{\infty} [|\nabla q|^2 - \beta |\partial q / \partial \tau|^2 - (1 - \sigma R)|q|^4 - 2pR|q|^2] d\eta d\zeta d\tau, \qquad (2)$$

$$L_\xi = \frac{\mathbf{e}_\xi}{2i} \iint \int_{-\infty}^{\infty} [\mathbf{r} \times (q^* \nabla q - q \nabla q^*)] d\eta d\zeta d\tau.$$

where $\nabla = \mathbf{e}_\eta \partial / \partial \eta + \mathbf{e}_\zeta \partial / \partial \zeta$, while $\mathbf{e}_\eta, \mathbf{e}_\zeta, \mathbf{e}_\xi$ are unit vectors in directions of $\eta, \zeta, \xi$ axes.

We search for light bullet solutions of Eq. (1) in the form $q = w(r, \tau) \exp(ib\xi)$, where $w$ is a real function and $b$ is the propagation constant. As in a usual linear lattice $(\sigma = 0)$, in the nonlinear lattice low-amplitude solitons, whose propagation constant $b$ is close to the lower cut-off for soliton existence $b_{\text{co}}$, strongly expand in both space and time [Figs. 1(a) and 1(b)]. Such bullets cover multiple lattice rings and acquire pronounced shape oscillations. Increasing the soliton amplitude results in light concentration near the central guiding lattice core. In this regime the effects of linear and nonlinear lattices are comparable and therefore the soliton maximum is located at $r = 0$.

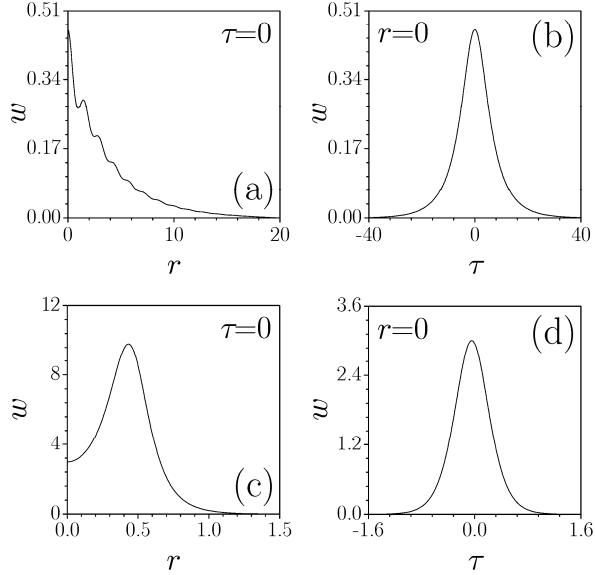

Fig. 1. (a),(c) Spatial field distributions at $\tau = 0$, and (b),(d) temporal field distributions at $r = 0$ in light bullets corresponding to $b = 0.02$ (top row) and $b = 18$ (bottom row) at $p = 3$, $\sigma = 0.9$.

As expected on physical grounds, a further growth of the soliton amplitude results in a large transformation of the spatial shape of the light bullets due to the fact that the nonlinear contribution to the refractive index starts dominating over the linear lattice and, therefore, light self-focused in the region where nonlinearity is stronger. This is accompanied by the development of ring-like spatial profiles [Figs. 1(c) and 1(d)], an effect that is most pronounced around the pulse peak $(\tau = 0)$. The temporal distributions do not change qualitatively and remain bell-shaped. Note that the shift of the intensity peak from the region where the linear index is maximal into the region with higher nonlinearity is possible also in 1D [39] and 2D [40] linear-nonlinear lattices. Such shape transformation causes non-monotonic dependencies of the energy $U$ on the propaga-tion constant $b$ [Fig. 2(a)]. Increasing the depth of the nonlinearity modulation results in a significant growth of the maximum possible energy that light bullets can carry. As $\sigma$ increases, the range of propagation constants and energies where the slope of $U(b)$ curve is positive continuously expands. The corresponding Hamiltonian-energy diagram exhibits a single cusp connecting the branches where $dU/db$ is positive [lower curve on $H(U)$ diagram] or negative [upper curve on $H(U)$ diagram] [Fig. 2(b)]. It is a well established fact that in lattices with uniform nonlinearity ($\sigma = 0$) the branches where $dU/db > 0$ correspond to stable solutions in accordance with the VK stability criterion. We found that when $\sigma \neq 0$ such criterion may fail.

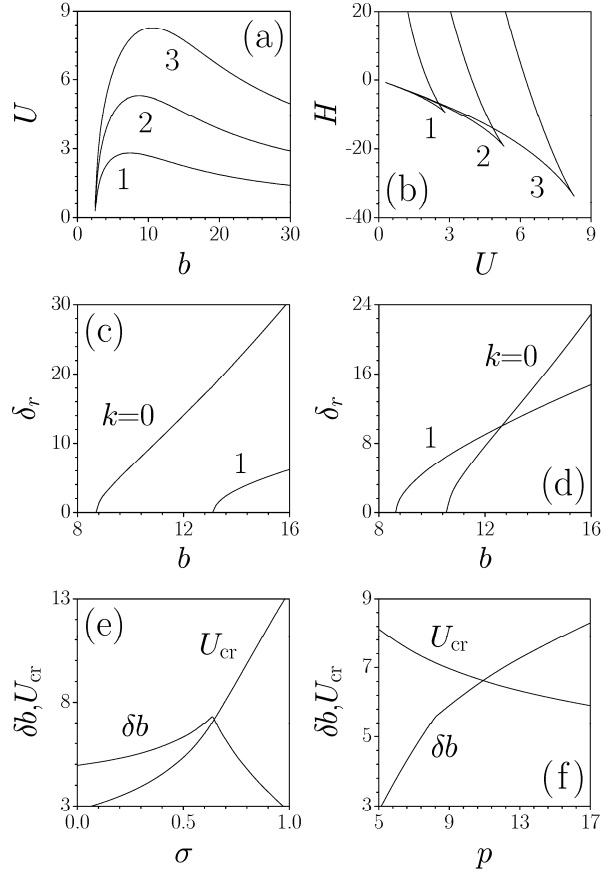

Fig. 2. (a) Energy versus propagation constant and (b) Hamiltonian versus energy for light bullets for $p = 12$ at $\sigma = 0.0$ (1), $0.5$ (2), and $0.7$ (3). Real part of perturbation growth rate versus $b$ for $p = 12$ at (c) $\sigma = 0.5$ and (d) $\sigma = 0.7$. Critical energy and $\delta b = b_{cr} - b_{co}$ (e) versus $\sigma$ at $p = 12$ and (f) versus $p$ at $\sigma = 0.6$.

To gain further insight into the light bullet stability, we conducted a detailed, rigorous linear stability analysis of the stationary solutions by looking at the evolution of perturbed solutions in the form $q = [w + u\exp(ik\phi + \delta\xi) + v^*\exp(-ik\phi + \delta^*\xi)]\exp(ib\xi)$, with $u, v$ being small perturbations, corresponding to azimuthal perturbation index $k$. Substitution of this expression into Eq. (1) and linearization around $w(r, \tau)$ yields the linear eigenvalue problem for the perturbations $u, v$, which may grow with a complex rate $\delta = \delta_r + i\delta_i$:

$$i\delta u = -\frac{1}{2}\left(\frac{\partial^2 u}{\partial r^2} + \frac{1}{r}\frac{\partial u}{\partial r} - \frac{k^2 u}{r^2} + \frac{\partial^2 u}{\partial \tau^2}\right) + bu - pRu - (1-\sigma R)w^2(2u+v),$$

$$i\delta v = +\frac{1}{2}\left(\frac{\partial^2 v}{\partial r^2} + \frac{1}{r}\frac{\partial v}{\partial r} - \frac{k^2 v}{r^2} + \frac{\partial^2 v}{\partial \tau^2}\right) - bv + pRv + (1-\sigma R)w^2(2v+u).$$
(3)

While in lattices with $\sigma = 0$, only perturbations with $k=0$ may destabilize the light bullet solutions, it turns out that in lattices with nonlinearity modulation such as the one addressed in this paper, the perturbation modes corresponding to $k=1$ may also become destructive, mainly because high-amplitude bullets tend to develop a ring-like structure in space already at moderate $\sigma$ values. Our findings indicate that the azimuthal instability associated with $k=1$ becomes especially pronounced when the depth of nonlinearity modulation exceeds a critical value $\sigma_{cr}$ that depends on $p$ (for example, at $p=12$ one gets $\sigma_{cr} \approx 0.65$).

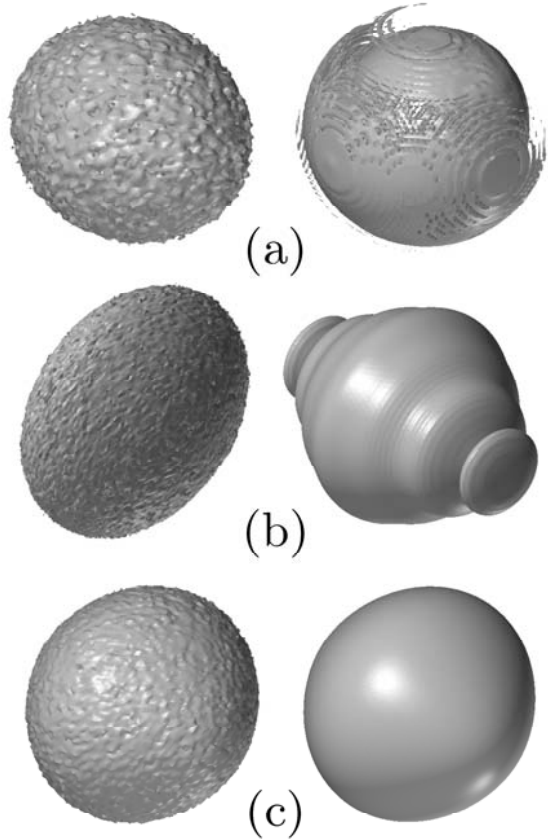

Fig. 3. Isosurface plots showing propagation dynamics of perturbed light bullets corresponding to (a) $b=10$, (b) $b=12$, and (c) $b=8$. Left column shows input patterns at $\xi=0$, while right column shows output patters at (a) $\xi=0.8$, (b) $\xi=1$, and (c) $\xi=300$. In all cases $p=12$, $\sigma=0.7$.

Typical dependencies of the perturbation growth rate on the propagation constant are shown in Figs. 2(c) and 2(d) for $\sigma < \sigma_{cr}$ and $\sigma > \sigma_{cr}$, respectively. One can see that at $\sigma < \sigma_{cr}$ the instability associated with $k=0$ appears at smaller $b$ values (exactly corresponding to the point where $dU/db$ changes its sign) than the instability corresponding to $k = 1$ and remains dominant for all $b$. The behavior is the opposite for $\sigma > \sigma_{cr}$ where the instability associated with $k = 1$ appears earlier than the instability corresponding to $k = 0$. Therefore, bullets in lattices with sufficiently strong modulation of nonlinearity may be unstable even in the region where $dU/db > 0$ (note that VK criterion does not hold starting exactly at the point where solitons become unstable with respect to the perturbation mode corresponding to $k = 1$). Figure 2(e) shows the width of the stability domain, defined as $\delta b = b_{cr} - b_{co}$ (here $b_{cr}$ is the propagation constant value at which bullet becomes unstable) versus nonlinearity modulation depth. One can see that the width of the stability domain in terms of propagation constant first increases with $\sigma$, acquires its maximal value at $\sigma = \sigma_{cr}$, and then starts decreasing. In contrast, the maximal possible energy $U_{cr} = U(b_{cr})$ of stable bullets is a monotonically increasing function of $\sigma$ [Fig. 2(e)]. Finally, we show in Fig. 2(f) how the width of the stability domain changes with the depth $p$ of linear lattice when $\sigma$ is kept fixed. We found that the width of stability domain monotonically increases with $p$, but the slope of the dependence $\delta b(p)$ changes abruptly at a certain critical lattice depth. Such result occurs because the upper boundary of the instability domain at small $p$ is determined by the perturbation corresponding to $k = 1$ (this instability comes into play at smaller $b$ values than that corresponding to $k = 0$), while with increasing $p$ the situation gradually reverses and the perturbation corresponding to $k = 0$ appears earlier. The change of the slope of $\delta b(p)$ dependence occurs exactly at the point where both types of instabilities appear simultaneously. Note also that the maximal energy carried by stable bullets monotonically decreases with increasing lattice depth $p$.

The above results of the linear stability analysis were confirmed by direct propagation of the stationary solutions, perturbed by initial additive random fluctuations. Illustrative results are shown in Fig. 3. High-amplitude bullets, which are predicted to be unstable only to perturbations with azimuthal index $k = 1$, gradually loose their radial symmetry upon propagation due to the development of azimuthal modulations accompanied by radiation emission [Fig. 3(a)]. Bullets that are affected by instabilities with $k = 0$ either experience diffraction [Fig. 3(b)] or abrupt contraction, depending on the particular shape of the initial perturbation. In contrast, bullets with moderate amplitudes and propagation constants falling into the stability domain, retain their input shapes upon propagation even in the presence of considerable input noise [Fig.3(c)].

Finally, it is worth mentioning that the Bessel linear-nonlinear lattices addressed here can also support light bullets with vorticity, i.e. higher-order soliton solutions of Eq. (1) that can be written in the form $q = w(r,\tau)\exp(im\phi)\exp(ib\xi)$. Because the intensity of such bullets carrying a phase singularity must vanish at $r \to 0$, zero-order Bessel lattices featuring a maximum at $r = 0$ are not suitable for the formation of such states. Instead, one may utilize higher-order Bessel lattices described by the function $R(r) = J_n[(2b_{lin})^{1/2}r]$, where $n = 1,2,...$ . Such lattices feature clearly pronounced central guiding rings, where vortex-soliton light bullets can form. As in the case of fundamental light bullets, we found that the competition between spatially modulated nonlinearity and linear lattice results in shape transformation of bullets carrying vorticity. A growth of the peak intensity is accompanied by an expulsion of bright vortex ring from the first ring of the linear lattice into the region between rings, where the nonlinearity is higher. As in the case of fundamental bullets, the $U(b)$ dependencies for bullets carrying vorticity are nonmonotonic. Our calculations indicate that all nonlinear bullets with vorticity decay via azimuthal instabilities. The properties of the fundamental light bullets supported by higher-order Bessel lattices are qualitatively similar to the properties of bullets in zero-order lattices described here.

In summary, we addressed light bullets supported by two-dimensional Bessel optical lattices with out-of-phase modulation of linear and nonlinear refractive indices. We found that the competition between out-of-phase linear and nonlinear lattices substantially modifies significantly the stability properties of light bullets. In particular, we showed that increasing the depth of the nonlinearity modulation results in a growth of the maximum energy carried by the bullets.

**Acknowledgements**

This work has been supported by the Hong Kong Baptist University and Hong Kong Research Grants Council, and by the Government of Spain through the Ramon-y-Cajal program.